\documentclass[aps,twocolumn,superscriptaddress,floatfix,showpacs,preprintnumbers,amsmath,amssymb]{revtex4}
\usepackage[colorlinks=true,citecolor=blue,linkcolor=blue]{hyperref}
\usepackage[usenames]{color}
\usepackage{txfonts}
\usepackage{dcolumn}
\usepackage{mathrsfs}
\usepackage{bm}
\usepackage{amsmath,amssymb,epsfig,float,graphics}

\begin{document}
\baselineskip=0.45 cm

\title{Analog cosmological particle generation in a superconducting circuit}

\author{Zehua Tian}
\email{zehuatian@126.com}
\affiliation{Institute of Theoretical Physics, University of Warsaw, Pasteura 5, 02-093 Warsaw, Poland}

\author{Jiliang Jing}
\email{jljing@hunn.edu.cn}
\affiliation{Department of Physics, Key Laboratory of Low Dimensional Quantum Structures and Quantum Control of Ministry of Education, 
Hunan Normal University, Changsha, Hunan 410081, P. R. China}
\affiliation{Synergetic Innovation Center for Quantum Effects and Applications, Hunan Normal University, 
Changsha, Hunan 410081, P. R. China}

\author{Andrzej Dragan}
\email{dragan@fuw.edu.pl}
\affiliation{Institute of Theoretical Physics, University of Warsaw, Pasteura 5, 02-093 Warsaw, Poland}

\begin{abstract}

We propose the use of a waveguidelike transmission line based on direct-current superconducting quantum interference devices (dc-SQUID)
and demonstrate that the node flux in this transmission line behaves in the same way as quantum fields in an expanding (or contracting) universe. 
We show how to detect the analog cosmological particle generation and analyze its feasibility with current circuit quantum electrodynamics (cQED) technology. 
Our setup in principle paves a new way for the exploration of analogue quantum gravitational effects.

\end{abstract}

\baselineskip=0.45 cm
\maketitle
\newpage
\section{Introduction} 

One of the most striking phenomena predicted by the quantum field theory in curved spacetime is the cosmological particle creation, i.e., particles could be spontaneously created out of the (virtual) quantum vacuum fluctuations as a consequence of the expansion of the Universe \cite{Birrell}. This mechanism is responsible for the generation of the seeds for cosmic structure formation, and thus plays a very important role in the past and future fate of our Universe. For this reason, recently there has been growing interest in theoretical studies involving this nontrivial quantum effect \cite{Gibbons, Greg, Ivette, Ball, Yasusada, Martin, Tian1, Tian2, Shahpoor, Wang, Helder}. 

However, cosmology in the early Universe is far away from everyday experience, because there has in fact ``only ever been one experiment, still running, and we are latecomers watching from the back"  \cite{Pickett}. Although there can be no truly experimental cosmology, it would be desirable to render the relevant phenomena accessible to an experimental investigation. In order to achieve this goal, a  promising way is to construct the ``analogous gravity"  system where the relevant features of quantum fields in curved spacetime can be reproduced analogously.  Along this line of reasoning, a lot of ``analogous gravity" experiments involving many nascent yet fast-growing fields, such as Bose-Einstein-Condensates \cite{Petr, Fischer, Jain, Angus, Carlos1, Carlos2, Clemens} and ion trap \cite{Alsing, Ralf, Nicolas, Christian}, have been proposed to observe the analogue cosmological particle creation.

On the other hand, as a promising candidate for future quantum information processing, circuit quantum electrodynamics (cQED) \cite{Wallraff} could offer a natural arena for testing fundamentals of quantum mechanics and implementing quantum field theory (QFT) concepts \cite{Nation1} due to its fantastic controllability and scalability. Based on this technology, a lot of theoretical proposals for analog circuit realizations including Hawking radiation \cite{Nation2}, traversable wormhole spacetime \cite{Sabin}, Fermion-Fermion scattering in quantum field theory \cite{Garcia}, and dynamical gauge fields \cite{Marcos, Mezzacapo}, have been demonstrated. Moreover, experimentally, the dynamic Casimir effect--the generation of particles out of the quantum vacuum fluctuation due to the motion of boundary conditions--has been observed in a coplanar waveguide (CPW) terminated by a SQUID \cite{Wilson}, or in a Josephson metamaterial \cite{Lahteenmaki}. These theoretical and experimental studies open a new avenue to construct the experimental platform for the research on the relativistic QFT \cite{ Nation1,Joel, Felicetti, Paulina, Maximilian} and relativistic quantum information \cite{Friis, Laura}.

Here, based on the cQED technology, we propose using a superconducting electrical circuits configuration based on micro-fabricated waveguides and dc-SQUID. We demonstrate the analogy between magnetic flux in this superconducting circuit and quantum fields in an expanding (or contracting) universe. We also analyze how to 
detect the analogue cosmological particle creation and discuss its experimental feasibility with current cQED technology.
Our technique provides a novel tool for quantum simulations of relativistic QFT and in principle enables the exploration of analogue quantum gravitational effects.


\section{Physical model}

\begin{figure}[ht]
\centering
\includegraphics[width=0.46\textwidth]{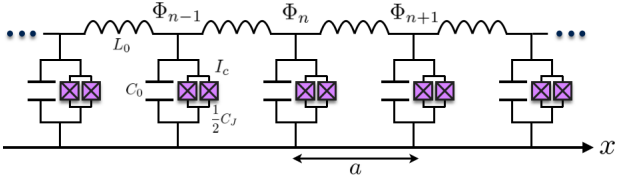}
\caption{(Color online) Circuit diagram for a coplanar waveguide-like transmission line. We assume the inductance for each inductor and the capacitance for 
each capacitor are $L_0$ and $C_0$, respectively. Each SQUID element is formed from two identical tunnel Josephson junctions with critical current $I_c$ and capacitance $\frac{1}{2}C_J$. The length of all the cells is constant and is equal to $a$. The circuit is characterized by the dynamical fluxes $\Phi_n$.}\label{fig1}
\end{figure}

As shown in Fig. \ref{fig1}, we consider a coplanar transmission line which is similar to the CPW introduced in Refs. \cite{Johansson}. However, here each capacitor in the circuit is parallel with an identical SQUID, which consists of two parallel identical Josephson junctions (JJs) with the critical current $I_c$ and capacitance $\frac{1}{2}C_J$. In this work, we assume that the geometric size of the SQUID loop is small enough such that the SQUID's self-inductance is negligible compared to its kinetic inductance. Under this condition, each SQUID behaves like a single JJ with effective junction capacitance $C_J$ and
tunable Josephson energy $E_J(\Phi^J_\text{ext})=2E_J\bigg|\cos\bigg(\pi\frac{\Phi^J_\text{ext}}{\Phi_0}\bigg)\bigg|$ \cite{Johansson}. Here $\Phi_0=h/2e$ is the magnetic flux quantum, $E_J=\frac{\Phi_0I_c}{2\pi}$ is the Josephson energy, and $\Phi^J_\text{ext}= BA_S$ is the flux dropping through the SQUID loop with effective area $A_S$, and the applied magnetic field $B$.

For the system under consideration, its corresponding circuit Lagrangian reads:
\begin{eqnarray}\label{Lagrangian-S}
\nonumber
\mathscr{L}&=&\sum^{N}_{n=1}\bigg[\frac{1}{2}C_0\big(\dot{\Phi}_n\big)^2	-\frac{(\Phi_{n+1}-\Phi_n)^2}{2L_0}+\frac{1}{2}C_{J}\big(\dot{\Phi}_n\big)^2
\\
&&+E_{J}(\Phi^{J}_\text{ext})\cos\bigg(2\pi\frac{\Phi_n}{\Phi_0}\bigg)\bigg],
\end{eqnarray}
where $\Phi_n$ is the node flux. 
We restrict ourselves to frequencies in the circuit far below the plasma frequency of the SQUID so that oscillations in the phase across the SQUID could satisfy $\frac{\Phi_n}{\Phi_0}\ll1$, and the SQUID is operated in the phase regime where $E_J(\Phi^J_\text{ext})\gg(2e)^2/2C_J$. The small amplitude condition, $\frac{\Phi_n}{\Phi_0}\ll1$, allows us to study the Lagrangian above by linearizing the Josephson terms, i.e., expanding it up to the second order in $\Phi^2_n$. Besides, we assume that the wavelength $\lambda$ for the flux is much longer than the dimensions of a single unit cell of the chain, i.e., $(a/\lambda\ll1)$. Then we can use the continuum approximation by replacing the discrete $n$ by a continuous position $x$ along the line and replacing the finite difference in the Lagrangian by their continuous counterparts to first order in $(a/\lambda)$, i.e., $\Phi_n-\Phi_{n-1}\approx\,a\frac{\partial\Phi}{\partial\,x}+o(a^2)$. Finally, in the linear and continuum 
limit, the Lagrangian in Eq. \eqref{Lagrangian-S} can be rewritten as:
\begin{eqnarray}\label{L2}
\mathscr{L}=\frac{C}{2}\int\,dx\bigg[\bigg(\frac{\partial{\Phi}}{\partial\,t}\bigg)^2-\frac{a^2}{L_0C}\bigg(\frac{\partial\Phi}{\partial\,x}\bigg)^2
-\frac{1}{L_{J}C}\Phi^2\bigg],
\end{eqnarray}
where $C=C_0+C_{J}$, and $L_J=\bigg(\frac{\Phi_0}{2\pi}\bigg)^2/E_J(\Phi^J_\text{ext})$ is the kinetic inductance of the equivalent single JJ. Through variation, we can obtain the equation of motion of the flux $\Phi$:
\begin{eqnarray}\label{field1}
\frac{\partial^2\Phi}{\partial\,t^2}-\frac{a^2}{L_0C}\frac{\partial^2\Phi}{\partial\,x^2}+\frac{1}{L_{J}C}\Phi=0.
\end{eqnarray}
The last term in Eq. \eqref{field1} comes from the SQUID which in fact provides a potential energy in Lagrangian \eqref{L2} for the oscillator system. It is controllable 
because we can control the inductance $L_J$ by adjusting the external magnetic flux $\Phi^J_\text{ext}$ threading the SQUID loop. This property thus can allow us to
implement the requested potential energy that is space- or time-dependent. Let us note that this term is very important in our work, which is the key to simulating 
the cosmological particle creation in the following analysis.

\section{Cosmological particle creation.}

Consider a real massive scalar field $\Phi$ in $(1+1)$ dimensional spacetime, its corresponding action
is of:
\begin{eqnarray}\label{Field-action}
\mathcal{A}=\frac{1}{2}\int\,d^2x\sqrt{-\mathrm{g}}[(\partial_\mu\Phi)\mathrm{g}^{\mu\nu}(\partial_\nu\Phi)-m^2\Phi^2],
\end{eqnarray}
where $\mathrm{g}=\det\{\mathrm{g}_{\mu\nu}\}$ and $\mathrm{g}^{\mu\nu}$  are respectively the determinant and inverse of the spacetime metric tensor $g_{\mu\nu}$. 
To see how, in practice, particle creation can occur in an expanding (or contracting) universe, we will consider a two-dimensional Friedman-Robertson-Walker
universe with line element:
\begin{eqnarray}\label{FRW-metric}
\text{d}s^2=d\eta^2-\mathfrak{a}^2(\eta)dr^2,
\end{eqnarray}
where the time-dependent scale parameter $\mathfrak{a}(\eta)$ corresponds to the cosmic expansion/contraction. Introducing the conformal time $t$ associated with $\eta$ via $\eta=\int^\eta\,d\eta^\prime=\int^t\mathfrak{a}(t^\prime)dt^\prime$, the FRW line element \eqref{FRW-metric} may be recast as \cite{Birrell}:
\begin{eqnarray}\label{C-metric}
\text{d}s^2=\mathfrak{a}^2(t)(dt^2-dr^2).
\end{eqnarray}
This form of the line element is manifestly conformal to Minkowski spacetime with the so-called conformal scale factor $\mathfrak{a}^2(t)$. Using \eqref{Field-action} and \eqref{C-metric}, we obtain the field equation:
\begin{eqnarray}\label{field-equation}
\frac{\partial^2\Phi}{\partial\,t^2}-\frac{\partial^2\Phi}{\partial\,r^2}+\mathfrak{a}^2(t)m^2\Phi=0.
\end{eqnarray}

After a normal-mode expansion, Eqs. \eqref{field1} and \eqref{field-equation} becomes $\ddot{\Phi}_{k^\prime}+\big(\frac{a^2{k^\prime}^2}{L_0C}+\frac{1}{L_{J}C}\big)\Phi_{k^\prime}=0$ and $\ddot{\Phi}_k+\big(k^2+\mathfrak{a}^2(t)m^2\big)\Phi_k=0$, respectively. Comparing them and identifying $r$ with $x$, we observe a strong similarity: 
the factor $\frac{a^2{k^\prime}^2}{L_0C}$ acts as the wave number of field, corresponding to $k^2$, and $\frac{1}{L_{J}C}$ directly corresponding to $\mathfrak{a}^2(t)m^2$ acts like the conformal scale factor. Let us note that by appropriately adjusting the external magnetic flux $\Phi^J_\text{ext}$ through the SQUID loop, we can control $L_J$ to simulate the behavior of the conformal scale factor $\mathfrak{a}^2(t)$. Thus, the analogous dynamic behavior of quantum fields in an expanding/contracting universe in principle could be implemented in our setup.

\begin{figure}[ht]
\centering
\includegraphics[width=0.46\textwidth]{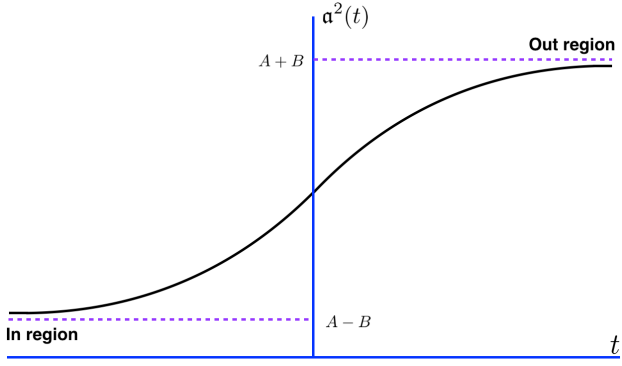}
\caption{(Color online) The conformal scale factor $\mathfrak{a}^2(t)=A+B\tanh\big(\rho\,t\big)$ represents an asymptotically static universe that undergoes a period of
smooth expansion. In region and out region correspond to $\mathfrak{a}^2(-\infty)=A-B$ and $\mathfrak{a}^2(\infty)=A+B$, respectively.}\label{fig2}
\end{figure}

Assuming that $\Phi^{\text{in}}_k$ and $\Phi^{\text{out}}_k$ are two particular solutions to Eq. \eqref{field-equation}, corresponding to solutions which converge, respectively, in the past $(t\rightarrow-\infty)$ and future $(t\rightarrow\infty)$ to plane waves. Physically, these two particular solutions represent a description of two different vacuum states, $|\text{in}\rangle$ and $|\text{out}\rangle$, in the two asymptotic regions where the function $\mathfrak{a}(t)$ becomes constant and the definition of particle state is unambiguous. In general, one has:
 \begin{eqnarray}\label{BLT}
\Phi^{\text{in}}_k=\alpha_k\Phi^{\text{out}}_k+\beta_k\Phi^{\text{out}\ast}_{-k},
\end{eqnarray}
with $\beta_k\neq0$. If we choose a toy, $\mathfrak{a}^2(t)=A+B\tanh\big(\rho\,t\big)$, which characterizes an asymptotically static universe that undergoes 
a period of smooth expansion shown in Fig. \ref{fig2},
then we can solve Eq. \eqref{field-equation} analytically. In this model, the inner products $\alpha_k$ and $\beta_k$ (known as Bogoliubov coefficients) are given exactly by the simple expressions:
\begin{eqnarray}
\nonumber
\alpha_k&=&\bigg(\frac{\omega_\text{out}}{\omega_\text{in}}\bigg)^{\frac{1}{2}}\frac{\Gamma(1-(i\omega_\text{in}/\rho))\Gamma(-i\omega_\text{out}/\rho)}{\Gamma(-i\omega_+/\rho)\Gamma(1-(i\omega_+/\rho))},
\\
\beta_k&=&\bigg(\frac{\omega_\text{out}}{\omega_\text{in}}\bigg)^{\frac{1}{2}}\frac{\Gamma(1-(i\omega_\text{in}/\rho))\Gamma(i\omega_\text{out}/\rho)}{\Gamma(i\omega_-/\rho)\Gamma(1+(i\omega_-/\rho))},
\end{eqnarray}
where $\Gamma$ is the Euler function, $\omega_\text{in}=[k^2+m^2(A-B)]^\frac{1}{2}$, $\omega_\text{out}=[k^2+m^2(A+B)]^\frac{1}{2}$, and
$\omega_\pm=\frac{1}{2}(\omega_\text{out}\pm\omega_\text{in})$. Obviously, the normalization condition $|\alpha_k|^2-|\beta_k|^2=1$ follows immediately. Let us note that although the initial vacuum state $|\text{in}\rangle$ containing no particles seen from all the inertial observers in the remote past ($t\rightarrow-\infty$), it is not the physical vacuum anymore when detected by the inertial observer in the out region ($t\rightarrow\infty$). The particle detectors there will register the presence of quanta. In mode $k$, the expected number of detected quanta is \cite{Birrell}:
\begin{eqnarray}\label{number-particles}
\langle\,\hat{n}_k\rangle=|\beta_k|^2=\frac{\sinh^2(\pi\omega_-/\rho)}{\sinh(\pi\omega_\text{in}/\rho)\sinh(\pi\omega_\text{out}/\rho)}.
\end{eqnarray}
This is the creation of particles into the mode $k$ as a consequence of the cosmic expansion/contraction.


\section{Experimental implementation}

To realize the proposed experiment, an additional conducting line is needed to produce the time varying external flux bias $\Phi^J_\text{ext}$, which, as discussed above, is used to modulate the SQUIDs for providing a time-dependent potential energy. More specifically, 
\begin{eqnarray}
\Phi^J_\text{ext}=\frac{\Phi_0}{\pi}\arccos\bigg[\frac{\Phi_0Cm^2\mathfrak{a}^2(t)}{4\pi I_c}\bigg].
\end{eqnarray}
This required magnetic flux can be achieved by corresponding current through the bias line. It is unavoidable that the current pulse dispersion in the bias line will affect the cosmological particle creation. However, such effects in principle could be weakened by appropriate choice of pulse shape and transmission line length. Additionally, to observe the generated particles due to the cosmic expansion/contraction, frequency-tunable, single-shot photon detection at the end of transmission line opposite to that of the bias pulse origin is required. Here we will assume a superconducting phase-qubit as our model detector and detect the microwave photon based on the recently proposed technologies \cite{Hofheinz, Romero, Chen, Oleksandr}. If the current pulses are repeatedly injected down the bias line, the predicted particles will be
created as a consequence of the cosmic expansion/contraction. Correspondingly, its spectrum can be detected by tuning the qubit resonant frequency. 

We want to point out that the emitted photon pairs are in fact entangled \cite{Ivette, Ball, Martin}, and this entanglement, by coincidence detection, could be demonstrated. Besides, such entanglement is essential and nontrivial as a result of cosmic expansion/contraction \cite{Eg}. It is needed in the outcome to establish that a photon is produced by the cosmology expansion/contraction rather than other irrelevant processes, such as ambient emission or capacitive coupling to the bias line. Let us note that properly engineering the transmission line could effectively reduce the background noise from the unwanted coupling.

\begin{figure}[ht]
\centering
\includegraphics[width=0.48\textwidth]{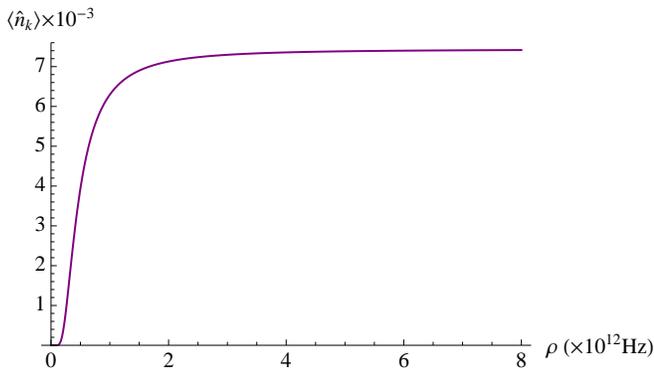}
\caption{(Color online) The expected number of detected quanta as a function of cosmic expansion-related parameter $\rho$. Here we assume the critical current  $I_c=1\,\mu\mathrm{A}$, and capacitance $C_J=0.5\,\mathrm{fF}$ for the JJ. The capacitance to ground is fixed as $C_0=0.1\,\mathrm{pF}$, and the inductance and the length of the single unit cell of our setup are assumed to be $L_0=0.25\,\mathrm{nH}$ and $a=15\,\mu\mathrm{m}$, respectively.}\label{fig3}
\end{figure}

We will chose the relevant parameters for each element similar to Refs. \cite{Nation2, Johansson, Castellanos} and estimate the cosmological particle generation in our setup.
For the JJ, we choose $I_c=1\,\mu\mathrm{A}$ and $C_J=0.5\,\mathrm{fF}$. Besides, the capacitance to ground is chosen as $C_0=0.1\,\mathrm{pF}$, the inductance and the length of the single unit cell of our setup are respectively assumed to be $L_0=0.25\,\mathrm{nH}$ and $a=15\,\mu\mathrm{m}$. In such case, we can have $\omega_\text{in}\approx0.21\times10^{12}\,\mathrm{Hz}$, and $\omega_\text{out}\approx0.25\times10^{12}\,\mathrm{Hz}$ by properly assuming the external magnetic flux, $\Phi^J_\text{ext}$, and the wave vector, $k^\prime$. In Fig. \ref{fig3}, we plot the number of created particles as a function of $\rho$ which is a positive real parameter controlling the rapidity of the expansion of universe. It shows that as $\rho$ is increased, more particles would be created. Besides, there is an asymptotic regime when $\rho\gg\omega_\text{out}$.

Let us discuss the number of created particles in the sudden approximation \cite{Carlos1}, i.e., $\rho\gg\omega_\text{out}$. Mathematically the ``sudden limit" consists of taking a step function for the scale factor, $\mathfrak{a}^2(t)=A-B+2B\,\Theta(t)$. Physically this means that one is considering that the change in $\mathfrak{a}(t)$ is driven more rapidly than the frequency band one is interested in. To implement the sudden approximation in experiment, we need to suddenly change the external magnetic fields from $\Phi^J_\text{ext}=\frac{\Phi_0}{\pi}\arccos\bigg[\frac{\Phi_0Cm^2}{4\pi I_c}(A-B)\bigg]$ to $\Phi^J_\text{ext}=\frac{\Phi_0}{\pi}\arccos\bigg[\frac{\Phi_0Cm^2}{4\pi I_c}(A+B)\bigg]$. For such case, we can get the expected number of detected quanta, $\langle\hat{n}_k\rangle=\frac{(\omega_\text{out}-\omega_\text{in})^2}{4\omega_\text{in}\omega_\text{out}}$, in mode $k$. With above parameters chosen, we find that in the ``sudden limit", this power per unit bandwidth in wave vector, $k$, range could give an energy comparable to a few $\mathrm{mK}$, which can be a factor of $10$ larger than the ambient temperature set by a dilution refrigerator. Therefore, this effect should be visible above the background noise.

\section{Discussions and Conclusions}

The parameters and pulse shapes above were chosen as an example that our setup is feasible, which should not be considered as the only available configuration. In fact, it is possible to improve and optimize these values in terms of both performance and fabrication of this proposal. Besides, we can also choose the conformal scale factor $\mathfrak{a}^2(t)$ as different functions to characterize different kinds of expansion/contraction of universe. It allows us to find a better physical model which could make our proposal more controllable and achievable in the future experiment. 

In summary, we have provided a recipe to build up an analog quantum simulator of Friedman-Robertson-Walker universe by means of a suitable strongly inhomogeneous external magnetic flux bias along a waveguide-like transmission line. The analogue quantum dynamics between the magnetic flux in this transmission line and the quantum fields in an expanding/contracting universe has been analyzed. Our results showed that the proposed device works in the quantum region and allows to observe the analogue cosmological particle generation.


\begin{acknowledgments}

We thank M. P. Blencowe for his valuable comments and inspiring discussions. Z. T. and A. D. thank the National Science Center, 
Sonata BIS Grant No. DEC-2012/07/E/ ST2/01402 for the financial support. J. J. is supported by the National Natural Science Foundation of China under
Grant No. 11475061.
\end{acknowledgments}



\begin{thebibliography}{99}
\bibitem{Eg}
This entanglement is the entanglement between the generated particles, i.e., between fields mode $k$ and $-k$ shown in Eq. \eqref{BLT}.
\end{thebibliography}
\end{document}